\documentclass[sigconf]{acmart}

\AtBeginDocument{%
  }

\copyrightyear{2025}
\acmYear{2025}
\setcopyright{cc}
\setcctype{by}
\acmConference[Websci '25]{Proceedings of the 17th ACM Web Science Conference 2025}{May 20--24, 2025}{New Brunswick, NJ, USA}
\acmBooktitle{Proceedings of the 17th ACM Web Science Conference 2025 (Websci '25), May 20--24, 2025, New Brunswick, NJ, USA}
\acmDOI{10.1145/3717867.3717917}
\acmISBN{979-8-4007-1483-2/2025/05}

\usepackage{booktabs}
\usepackage{graphicx}
\usepackage{subfigure}
\usepackage{afterpage}
\begin{document}

\title[The Role of Organizations in Networked Mobilization]{The Role of Organizations in Networked Mobilization: Examining the 2011 Chilean Student Movement Through The Logic of Connective Action}

\author{Diego G\'omez-Zar\'a}
\orcid{0000-0002-4609-6293}
\email{dgomezara@nd.edu}
\affiliation{%
  \institution{University of Notre Dame}
  \city{Notre Dame}
  \state{Indiana}
  \country{USA}
}

\author{Carolina P\'erez-Arredondo}
\email{carolina.pereza@uoh.cl}
\affiliation{%
  \institution{Universidad de O'Higgins}
  \city{Rancagua}
  \country{Chile}}

\author{Denis Parra}
\email{dparras@uc.cl}
\affiliation{%
  \institution{P. Universidad Cat\'olica de Chile}
  \city{Santiago}
  \country{Chile}}

\renewcommand{\shortauthors}{Gomez-Zara et al.}

\begin{abstract}
  This study examines the communication mechanisms that shape the formation of digitally-enabled mobilization networks. Informed by the logic of connective action, we postulate that the emergence of networks enabled by organizations and individuals is differentiated by network and framing mechanisms. From a case comparison within two mobilization networks---one crowd-enabled and one organizationally-enabled---of the 2011 Chilean student movement, we analyze their network structures and users' communication roles. We found that organizationally-enabled networks are likely to form from hierarchical cascades and crowd-enabled networks are likely to form from triadic closure mechanisms. Moreover, we found that organizations are essential for both kinds of networks: compared to individuals, organizations spread more messages among unconnected users, and organizations' messages are more likely to be spread. We discuss our findings in light of the network mechanisms and participation of organizations and influential users.
\end{abstract}




\maketitle

\section{Introduction}
Over the past few years, citizens have used social media platforms to start debates and mobilizations. Compared with decades ago, when formal and central organizations were mainly responsible for social protests, individuals have started mobilizations by communicating, organizing, and coordinating among decentralized networks on online platforms \cite{Bennett2013,Mirbabaie2021}. As a result, people from different spectrums can engage with these contemporary mobilizations without the presence of organizations or by being members of them. While several scholars have posited that organizational structures seem less necessary than they used to be, organizations' role during the formation of a digitally-enabled mobilization has not been explored in depth. Are organizations still relevant for the emergence of these mediated mobilization networks, whether individuals initiate them or not? \cite{Halupka2016,Karpf2016}. More importantly, little attention has been paid to how the initial sequence of users' interactions can launch a massive mobilization on these social media platforms \cite{GonzalezBailon2016}. 

This paper addresses these concerns by examining the communication roles and mechanisms that influence the formation of digitally-enabled mobilization networks. Informed by the logic of connective action \cite{Bennett2013}, we argue that organizations' and individuals' participation allows these mobilizations to form based on different network structures and framing strategies. We use the 2011 Chilean student movement as an example to illustrate our approach and its application. This historical case study represents approximations of the ideal connective action networks introduced by Bennett and Segerberg (\citeyear{Bennett2013}), where organizations and individuals collectively demanded new funding mechanisms for higher education. 

This work makes three contributions. First, it emphasizes that networks enabled by organizations are likely to propagate through hierarchical cascades (i.e., unconnected information chains from a few sources) due to the high deployment by organizations and leaders. Second, it shows that networks enabled by individuals are likely to propagate through triadic closure mechanisms (i.e., information shared through multiple and direct contacts) due to the high level of adaptability that the messages allow, which can be easily adopted and shared when they come from direct contacts. Third, it posits that organizations are still relevant for mobilizations enabled in social media networks. Even though users can promote messages without centralized structures, we contend that organizations, their leaders, and influential users are more likely to act as informational hubs and retransmit their messages to other activists. As a result, organizations are more likely than individuals to spread messages promoted by less-connected users to bigger audiences \cite{Stier2018}. Drawing on two Twitter mobilization networks extracted from hashtags, we examine how the communication mechanisms and roles of organizations and individuals shape their formation and diffusion.

\section{Literature Review}
\subsection{The Logic of Connective Action}
Since the late 2000s, inter-communicational processes among people have become easily and dynamically shared, fostering reliance on online technologies to coordinate actions, resist authority, bypass mass media, and amplify demands to broader audiences \cite{Castells2011}. In this context, the logic of connective action has shaped digital communication as a basic form of collective organization \cite{Bennett2013}. 

The logic of connective action accounts for individual action within the pursuit of collective goals, which can overcome hierarchical structures and organizational entities, and articulate personalized communication mechanisms to get individuals involved in a social cause. Unlike collective action, it posits that social action emerges from personal action frames shared by digital media users, who distribute messages to others using their own or others' terms \cite{Bennett2013}. These frames are both symbolically inclusive because they appeal to different personal reasons for contesting a situation that has to be changed and technologically open since individuals can transform these frames into several modes and share them through different channels. The high personalization of these frames through creative and captivating resources (e.g., images, videos, memes, and multimedia) enables users to easily engage with the movement without the need to participate in demanding activities or yield to a determined political identity. Therefore, social media facilitates the political involvement of people who are less identified with politics \cite{Vromen2015}. The focus relies on individuals' mobilization of massive protests without depending on organizational structures or top-down diffusion strategies \cite{Dessewffy2016}, where participation in social media networks (e.g., liking a Facebook post) becomes an act of personal expression validated, shared, and amplified by many others \cite{Boler2014}.

Connective action distinguishes three types of large-scale action networks that emerge from users' interactions. First, \textit{crowd-enabled networks} allow individuals and dispersed groups to employ personal action frames that travel through layers of technological platforms. Second, \textit{organizationally-enabled networks} allow organizations to seek loose network ties with coalition partners and leaders to activate individual-level networks by deploying personalized communication mechanisms. Finally, \textit{organizationally-brokered networks} encompass traditional collective action dynamics, that is, they manage the framing of action, provide leadership and resources, and try to broker differences that have to be bridged to form cohesive networks.

Extensive research has examined different mobilization networks worldwide using this theoretical framework. For example, Shanin et al. \cite{Shahin2024} examined how cross-cultural differences led to different framing strategies for the Black Lives Matter movement, showing that intra-cultural tensions within countries shape the construction of these connective frames. Mirbabaie et al. \cite{Mirbabaie2021} analyzed the \#metoo movement and identified two key archetypes in these networks: starters, who initiate and share messages at the beginning of a movement, and maintainers, who disseminate content and information across the mobilization network. Lastly, Tang \cite{Tang2023} introduced the concept of ``issue influencers,'' referring to users shaping mobilization networks around contentious political issues. 

In this paper, we focus on how these digitally-enabled mobilization networks form by drawing upon concepts from the logic of connective action. While previous studies have shown that digitally-enabled social movements' expansion relies on the participation of highly connected users, such as organizations, leaders, influential users, and celebrities \cite{GonzalezBailon2011,Tremayne2014}, little attention has been paid to how the initial sequence of users' interactions shape the formation of these networks and how the personal action frames will spread. We build upon the individuals' and organizations' communication mechanisms and roles to comprehend how these mobilization networks form. Moreover, we consider how network structural signatures influence the spread of messages, which thousands of protesters can ultimately adopt.

\begin{figure}[!htb]
    \centering
    \includegraphics[width=1\linewidth]{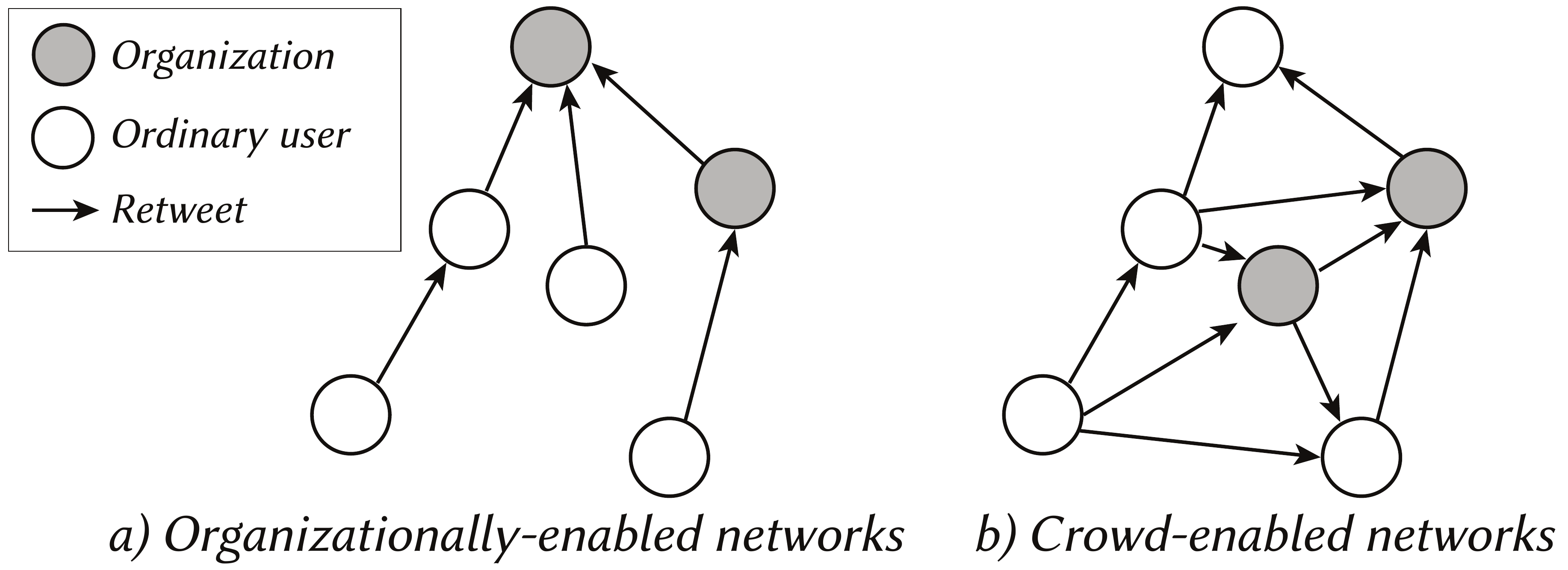}
    \caption{The structures of organizationally-enabled and crowd-enabled networks. Both networks have the same number of users, but the number of edges is different as users retweeted in different ways. Organizations are represented as grey circles, ordinary users are blank circles, and links are retweets.}
    \Description{Structures of organizationally-enabled and crowd-enabled networks. Both networks have the same number of users, but the number of edges is different as users retweeted in different ways. Organizations are represented as grey circles, ordinary users are blank circles, and links are retweets.}
    \label{fig:structures}
\end{figure}

\subsection{Structural signatures in digitally-enabled mobilization networks}
We expect that protest networks enabled by organizations will likely form through hierarchical cascades, that is, through information chains from a few sources (Figure \ref{fig:structures}a). First, organizations are more likely to activate their sparse networks to make their frames more visible. Compared to individuals, organizations are more likely to reach out to more digital media users because they strategically manage multiple resources to keep control of the frames and mobilize activists \cite{Diani2003,McCarthy1977}. Organizations' networks are less constrained than individuals, outreaching participants who are not necessarily connected and promoting frames through their online communication channels, leaders' accounts, websites, and releasing recruitment materials. Like in collective action, organizations will take advantage of their leaders' popularity and charisma to spread information in digital media \cite{Morris2004}. Similarly, organizations spread their messages to bigger audiences compared to those that individuals can outreach because of their high control of content, strong organizational coordination, and professional social media teams \cite{Gerbaudo2017,Pavan2019}. Therefore, organizations are more likely to become informational hubs among activists and spread information to different users who are not connected \cite{BarzilaiNahon2008}.

Second, organizations are likely to keep control of the messages' action frames because they manage core framing tasks and define specific frames to trigger action in their adherents. For these to succeed, Snow and colleagues \cite{Snow2018} argue that organizations focus on overcoming three main challenges to achieve resonance among the public: the identification of a) the problem (i.e., diagnostic framing), b) possible solutions (i.e., prognostic framing), and c) reasons to motivate adherents to act and join a specific cause (i.e., motivational framing). These frames are shaped and informed by various discursive opportunities and fields of action to achieve resonance \cite{McDonnell2017}. Most of these framing processes emerge at the organizational level to purposefully seek social change and galvanize the public \cite{Klandermans1984}. Organizations will promote messages and action frames to trigger specific discourses and social practices from their supporters, without considering how activists and supporters might appropriate their messages in line with their individual and collective identities. When their strategies are not perceived as successful, organizations reframe the narratives to reset their original strategies. Hence, the communication and replication of framing strategies carried out by organizations will reflect hierarchal cascade diffusion with key users in charge of its dissemination. Therefore, we postulate that:
\begin{quote}
\textit{H1: Organizationally-enabled networks are likely to form through hierarchical cascades than by chance.}
\end{quote}

In contrast, we expect that networks enabled by individuals are more likely to follow triadic closure mechanisms, that is, information will spread from people who share common connections (Figure \ref{fig:structures}b). Previous research on digital media has shown two main reasons why individuals are more likely to get information from the content close contacts and influential users share. First, individuals' networks are smaller and less strategic than organizations' networks since individuals are more likely to connect with others based on prior social relationships and their particular interests \cite{Bakshy2012,Contractor2006}. Individuals are also more likely to reach out to both strong and weak ties to promote social causes \cite{Cammaerts2013,Farrow2011} and share information with their contacts and, through them, the contacts of their contacts. Thus, people's willingness to adopt and share a personal action frame in digital media depends on the information spread in their personal networks and their need to strengthen their collective identities (Nekmat et al., 2015). Information that spreads through closed triads is associated with high levels of trust among individuals, reaching out to like-minded people, and exercising social influence \cite{Carullo2015}. Second, personal action frames created by individuals hinder ideological and definitional unanimity in their networks because they are designed to flexibly fit into each individual identity while addressing the formation and maintenance of collective identities. Ultimately, framing operates at the dyadic level, where personal discourses will be shared and re-personalized according to the similarity and identification between individuals. Therefore, we posit that:
\begin{quote}
    \textit{H2: Crowd-enabled networks are likely to form through triadic closure mechanisms than by chance.}	
\end{quote}

\subsection{Individuals versus organizations}
Although previous studies extensively acknowledge the efforts of individuals and organizations through their published content, technological use, and coalitions, these contributions are often described in broad, general terms, ignoring how these actors interact with each other in mobilization networks \cite{Earl2015}. Recent studies claim that both entities share each other's messages and that individuals may perform organizational strategies to coordinate and keep track of their actions \cite{Donovan2018,vonBulow2019}. In this shared digital ecosystem, we posit that organizations are more likely than individuals to promote messages in both kinds of networks since organizations are encouraged to gain visibility of their causes on digital media \cite{Leonardi2014}. Organizations also see digital media as public spaces to achieve connectivity and disseminate content collaboratively and asynchronously among potential supporters \cite{Gerbaudo2017,Lovejoy2012}. 

On the one hand, considering organizations want to influence the socio-political agenda, share information, and mobilize social media users \cite{Karpf2016}, their actions will become more prominent than those made by individuals when a network is organizationally-enabled. On the other hand, organizations will be more likely than individuals to share messages in crowd-enabled networks in order to monitor and control their political agendas. Moreover, individuals gain information and socialize with others, but most of them are not involved with diffusion and promotion on a daily basis \cite{Hsiao2018}. In contrast, organizations are more likely to gain knowledge and tactics and operationalize them in their activities \cite{Pavan2019}. Thus, organizations' participation is more likely to be consistent and strategic than the individuals', which requires a focus on news routines, media, and other stakeholders \cite{Earl2015}. We posit that:

\begin{quote}
    \textit{H3: Organizations are more likely to spread messages than individuals when mobilization networks form.}
\end{quote}

Finally, we seek to understand which users are more likely to be information sources as information is more likely to be shared through organizations than individuals in both kinds of networks. First, we argue that organizations are more likely to get followers who are looking for information related to mobilization efforts than individuals, thus acting as references for social change purposes \cite{Diani2003}. Consequently, potential supporters are more likely to mobilize around an issue that comes from organizational structures rather than unfamiliar groups \cite{KellyGarrett2006}. Second, organizations use social media to promote their news and causes, whereas individuals use social media mostly for interpersonal communication \cite{Lovejoy2012}. We expect that organizations will collect and distribute information from several sources, including individuals, to nurture collective efforts. Lastly, while information can come from less-connected or isolated users who are witnessing events or crafting the frames, the messages' visibility is lower because these users cannot outreach the rest of the network \cite{Barbera2015}. Organizations, as well as influential users, can make these messages more visible by spreading them to the rest of the network, becoming informational hubs of these networks \cite{Bakshy2011}. Considering these points, we posit that:

\begin{quote}
    \textit{H4: Messages published by organizations are more likely to be spread than messages published by individuals when mobilization networks form.}
\end{quote}

\section{The Case Study: The 2011 Chilean Student Movement}
The 2011 Chilean student movement gathered international attention through a wave of protests led by university students during the first right-wing government since the return of democracy in 1990. Although it has been more than ten years since this movement started, we decided to examine it for the following reasons. First, these protests were part of a global wave of social movements, including the Arab Spring and Occupy Wall Street. Yet, they have not been sufficiently examined in the literature compared to other global movements \cite{larrabure20152011,Bellei2014}. Studying this movement will help contextualize Chile's role within this broader historical moment of global unrest. Second, the 2011 movement laid the groundwork for the massive social protests in 2019 \cite{guzman2023power,gonzalez20202019}, which ultimately led to Chile's elections to rewrite its constitution. Third, many leaders of this movement were elected as the national government in 2022, with Gabriel Boric as the president. As such, studying this movement can contribute to our understanding of global mobilizations enabled by online platforms.  

The 2011 Chilean student movement had the participation of formal, annually elected student unions and emergent organizations. Each student union had a clear leadership structure, starting from the president and vice presidents to the inclusion of secretaries, which were voted on at the end of each academic year. The winners become the new representatives of each university and command the national student union called CONFECH (``Confederación de Estudiantes de Chile''), which sets this social movement's agenda \cite{Bellei2014}. Emergent organizations also participated and contributed to the organization of these protests. Students and citizens created their own civil groups and were represented under existing political parties' umbrellas. The novelty of this highly diverse movement is that its members and adherents correspond to minors and young adults, who are usually discarded as legitimate political actors under an adult-centric narrative \cite{PerezArredondo2019}. Through innovative protest repertoires, the movement influenced the political agenda and triggered social change through various discursive strategies that positioned them as legitimate political agents, thus reframing hegemonic discourses that sought their criminalization. In this process, Twitter was a crucial social media network for developing and maintaining this movement and informing and organizing events that brought students together for the same cause \cite{Valenzuela2012}. By using this platform, students actively sought to position themselves as political actors by openly challenging representations of themselves, their collectives, and the concept of national youth. 

To test our hypotheses, we studied one crowd-enabled network and an organizationally-enabled network that formed on Twitter. After examining hashtags that were trending topics on Twitter, we identified the following two hashtags that represented ideal approximations of both connective networks. 

\subsection{\#FuerzaEstudiantes: a crowd-enabled network}
By June 2011, students had occupied more than 100 high schools across the country to protest the commodification of education. Despite the government's repressive actions, the students' popular support continued to rise nationwide. In this context, the hashtag \#FuerzaEstudiantes (\#BeStrongStudents) emerged to rally demonstrators, share their achievements, and encourage other students to join their cause. The hashtag reached its peak in August 2011, after the brutal retaliation against mobilized students led by the police and authorized by the government \cite{PerezArredondo2012}. The hashtag foregrounded the various collective identities comprising and supporting the student movement, in which individual Twitter users aligned with the counter-hegemonic discourses constructed by and through students' protests and social practices, serving two different purposes: to reject the criminalizing framing with which the government and the media portrayed and delegitimized their actions, and to legitimate the need of a structural change to stop the commodification of peoples' right.  

\subsection{\#YoMarchoEl28: an organizationally-enabled network.}
A year later, CONFECH's leaders called for a strike on June 28th, 2012, and attempted to recreate the wave of support generated by students by individualizing the hashtag. The result, \#YoMarchoEl28 (``\#IMarchOnThe28'') aimed to exercise political pressure so the government met their demands. Rather than foregrounding the collective identities of the movement, the CONFECH attempted an individualization to highlight individuals' agencies. The goal seemed to appropriate the collective demands and channel them through individuals, thus departing from the homogenization strategies deployed by hegemonic discourses to delegitimize their claims. The protest took place in Santiago's main avenue, and more than 100,000 activists participated. 

\section{Methodology}
To study these two mobilization networks, we collected tweets containing these hashtags using the Twitter Academic API\footnote{The data was collected before the shutdown of the API.}. We used retweets (i.e., sharing someone else's tweet) to analyze how the information spread among users. Retweets were a traditional measure to assess influence and outreach on social media platforms during the 2010s \cite{Bakshy2011,boyd2010tweet}. Initial versions of Twitter indicated retweets by adding ``RT'' at the beginning of the tweet. Therefore, we followed a common practice and used the following keywords in the queries to identify retweets: ``RT @,'' ``MT @,'' and ``retweet @'' \cite{boyd2010tweet}.

\begin{figure*}[!htb]
    \centering
    \includegraphics[width=1\linewidth]{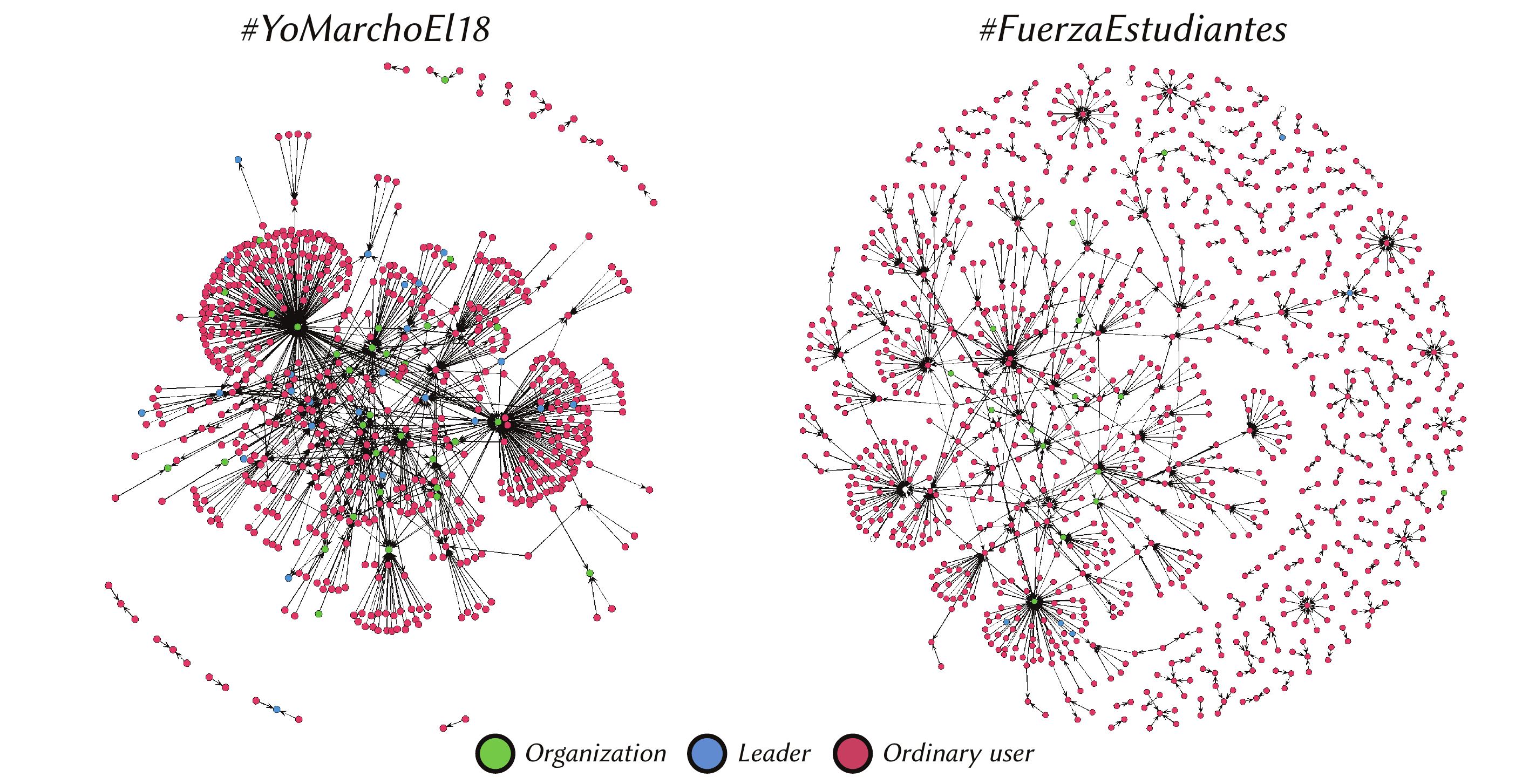}
    \caption{On the left, the organizationally-enabled network (\#YoMarchoEl18), and the crowd-enabled network (\#FuerzaEstudiantes) on the right. Red nodes represent ordinary users, green nodes represent organizations, and blue nodes represent leaders.}
    \label{fig:networks}
\end{figure*}

To compare the structure of these networks equally, we looked for the first 1,000 retweets of each hashtag on Twitter. We selected this threshold for several reasons: (a) to analyze the networks in the earlier stages and initial interactions, (b) to maintain a consistent basis for structural comparison during their formation, (c) to capture a sufficient number of interactions for meaningful network analysis using statistical modeling, and (d) to compare network growth and dynamics independent of overall activity over time. An alternative approach was to compare these networks over a fixed time span. However, their diffusion rates varied significantly: while \#FuerzaEstudiantes took 57 days to reach 1,000 retweets, \#YoMarchoEl28 achieved the same number in just eight days. 

We reviewed that the retweets were related to the movement and were not a consequence of ``hashtag hijacking.'' We then extracted the original tweet's author and the user who retweeted it to build retweet networks, one for each hashtag, adding users as nodes. These networks were directed since links go from the user who retweeted to the user who posted the message (Figure \ref{fig:networks}). Table \ref{tab:descriptives} presents the descriptive network statistics.

\subsection{Coding Twitter Accounts}
Two coders manually classified the Twitter accounts that participated in these networks based on whether they represented:
\begin{itemize}
    \item \textit{Ordinary users}: Twitter accounts of grassroots and individuals who were not directly related to a student organization.
    \item \textit{Student organizations}: Formal and emergent groups including student associations and college political parties.
    \item \textit{Student organization leaders}: Individuals who served as presidents, vice presidents, and secretaries of student organizations.
    \item \textit{Influential users}: Accounts from celebrities, politicians, and journalists found in the dataset.
\end{itemize}

The coders classified the accounts using the public information available on their Twitter profiles and reviewed student organizations' websites, news media articles, and political parties' websites during the coding process to verify these accounts. The coders achieved high consistency (Krippendorff $\alpha=0.85$) and resolved their differences by including the first author as the tie-breaker. Most importantly, we did not find any news media or bots as authors of these tweets. Moreover, we did not find influential users in the organizationally-enabled network within the first 1,000 tweets. Table \ref{tab:descriptives} shows the final users' classification, and Table \ref{tab:degree} their respective degree distributions. 

\subsection{Network analysis}
We used Exponential Random Graph Models (ERGM) to identify the individual, dyadic, and structural variables that best explain the motivations behind users' retweets. ERGM provides an appropriate statistical analytic methodology to test multi-theoretical multilevel network hypotheses \cite{Robins2007,Contractor2006}. It estimates the likelihood of the observed network structures emerging from all possible network configurations generated based on certain hypothesized self-organizing principles. We used ERGMs to model the retweet network as a function of individual-level variables, dyadic variables, and endogenous network structures as a whole. 

In these ERGMs, the dependent variable is the whole retweet network established by the users as one observation, and the independent variables are the network structural signatures and users' attributes. ERGM uses Maximum Likelihood Estimation (MLE) to estimate the network statistics' coefficients. Positive and significant coefficients indicate that the corresponding independent variable is more likely to influence users' sharing behavior than random chance, and negative and significant coefficients indicate that the independent variable is less likely to result in a user's retweet than chance alone. The effect size of one additional independent variable count is measured by the odds ratio (OR), which equals the exponential function of the corresponding coefficient (e.g., $e^\beta$). We formulated an ERGM for each network to test the structural signatures and users' roles. Figure \ref{fig:ergm-terms} shows the following independent variables, control variables, and their visual representations.

Once the ERGMs' coefficients were estimated, we tested whether these simulated networks fit within the observed network. To measure this fit, we used the simulation-based Goodness of Fit (GoF) test, which compares the characteristics of generated networks to the statistics of the observed network. We used R 4.4.0 \cite{RCoreTeam2019} and the package \texttt{statnet} \cite{Handcock2008} to develop these ERGMs. The script is available at OSF.io for revision\footnote{\href{https://osf.io/7wyg9/?view\_only=64794555dbca4b33aab39bfe60c90b04}{https://osf.io/7wyg9/?\=64794555dbca4b33aab39bfe60c90b04}}. 

\begin{table}[!ht]
\centering
\resizebox{\columnwidth}{!}{%
\begin{tabular}{@{}lcc@{}}
\toprule
 & \textbf{Crowd-enabled} & \textbf{Organizationally-enabled} \\ \midrule
Edges & 1,000 & 1,000 \\
Users & 1,026 & 684 \\
\textit{Number of accounts} &  &  \\
Organizations & 15 & 32 \\
Leaders & 5 & 24 \\
Ordinary users & 1,000 & 628 \\
Influential users & 6 & N/A \\
Centralization (degree) & 0.033 & 0.193 \\
Max. indegree & 65 & 242 \\
Max. outdegree & 10 & 34 \\ \bottomrule
\end{tabular}%
}
\caption{Descriptive statistics for the retweet networks}
\label{tab:descriptives}
\end{table}

\subsection{Independent variables}
\paragraph{Hierarchical cascades} To test H1, we included a term that accounts for the users' tendency to get information through cascades. From the ergm package, we added the directed geometrically weighted non-edgewise shared partners term (\texttt{dgwnsp}) using the ``outgoing two-path'' (OTP) specification. This term measures the extent to which messages from unconnected users pass through third users. A positive (negative) and significant term indicates that the frame is more (less) likely to spread through cascades than would be expected by chance. 

\paragraph{Triadic closure.} To test H2, we included a triadic term to account for the users' tendency to get the information from the original sources and their direct contacts. We included the directed geometrically weighted edgewise shared partners term (\texttt{dgwesp}) using the ``outgoing two-path'' (OTP) specification. Given a user \textit{i} who retweeted user \textit{j}, this term measures the extent to which user \textit{i} is also going to retweet other users who have already retweeted \textit{j}. A positive (negative) and significant term means that users are more (less) likely to spread the frame from the sources and direct contacts than chance.

\paragraph{Organizations' activity} To test H3, we added a categorical covariate to measure whether organizations retweeted more or less than ordinary users. A positive (negative) and significant term means organizations were more (less) likely to retweet than ordinary users. 

\paragraph{Organizations' popularity} To test H4, we added a categorical covariate to measure whether organizations were more likely to be retweeted than ordinary users. A positive (negative) and significant term shows that messages published by organizations were more (less) likely to be retweeted by other users.

\subsection{Control variables}
\paragraph{Density} This structural term represents the number of retweets in the network. A positive (negative) and significant term represents an active network with more (less) retweets than expected by chance. 

\begin{table}[!ht]
\centering
\resizebox{\columnwidth}{!}{%
\begin{tabular}{@{}lcc@{}}
\toprule
 & \textbf{Crowd-enabled} & \textbf{Organizationally-enabled} \\ \midrule
Organizations’ indegree & 7.33 (16.50) & 18.99 (49.39) \\
Organizations’ outdegree & 2.13 (2.50) & 4.44 (7.28) \\
Leaders’ indegree & 1.60 (3.58) & 3.25 (5.22) \\
Leaders’ outdegree & 0.80 (0.44) & 2.20 (2.20) \\
Influential users’ indegree & 8.33 (19.43) & N/A \\
Influential users’ outdegree & 0.66 (0.51) & N/A \\
Ordinary users’ indegree & 0.83 (2.67) & 0.50 (2.26) \\
Ordinary users’ outdegree & 0.96 (0.82) & 1.28 (1.15) \\ \bottomrule
\end{tabular}%
}
\caption{Users’ average degree in the retweet networks. Standard deviation in parenthesis}
\label{tab:degree}
\end{table}

\paragraph{Users' popularity} This structural term checks how centralized the retweets are on certain Twitter accounts. We included the geometrically weighted indegree term (\texttt{gwidegree}) from the \texttt{ergm} package, which estimates how concentrated the retweets are among certain users. A positive and significant term indicates a less centralized network where most users were retweeted, whereas a significant negative term indicates a skewed network where only a few users were retweeted \cite{Hunter2007}.

\paragraph{Users' activity} This term checks the extent to which retweets were made by certain users. We included the geometrically weighted outdegree term (\texttt{gwodegree}), which estimates how concentrated out-links are among certain users. Like the previous term, a positive and significant term indicates a less centralized network where most users retweeted, whereas a significant negative term indicates that certain users are responsible for most of the retweets. 

\paragraph{Users' number of followers} We analyzed whether the number of followers affected the number of retweets made or received by each user. We specified this attribute according to its initial users' number of followers when they first retweeted using the hashtag. We separated this covariate effect into sender and receiver effects. For the sender term, a positive (negative) and significant term means that users with many followers are more (less) likely to retweet. In the case of receiver effects, a positive (negative) and significant term shows that users with many followers are more (less) likely to be retweeted by other users.

\paragraph{Leaders and influential users' activity} We added a categorical covariate to measure whether leaders (influential users) were retweeted more or less than ordinary users. A positive and significant term means leaders (influential users) were more likely to retweet than ordinary users. 

\paragraph{Leaders and influential users' popularity} We added a categorical covariate to measure whether leaders (influential users) were more likely to be retweeted than ordinary users. A positive and significant term shows that messages published by leaders (influential users) were more likely to be retweeted by other users. 

\section{Results}
Table \ref{tab:results} presents the ERGMs' results predicting the likelihood of retweets among users in the crowd-enabled and organizationally-enabled networks. The GoF tests determined that the observed networks' statistics were well explained by the ERGMs, lying within 95\% of the confidence interval (See Appendix). 

H1 states that organizationally-enabled networks are more likely to form through hierarchical cascades. The hierarchical cascade term of the organizationally-enabled network's ERGM had a positive and significant effect ($\beta = 0.056, OR = 1.05, p< 0.01$). Messages passing through cascades were 1.05 times more likely to occur than chance. Thus, H1 is supported.

\begin{table}[!htb]
\centering
\resizebox{\columnwidth}{!}{%
\begin{tabular}{@{}lcc@{}}
\toprule
\textbf{Estimates} & \textbf{Crowd-enabled} & \textbf{Org-enabled} \\ \midrule
\textbf{Network signatures} &  &  \\
Density & -7.834 (0.21)*** & -5.583 (0.181)*** \\
Hierarchical cascades (H1) & 0.001 (0.03) & 0.056 (0.004)*** \\
Triadic closure (H2) & 2.722 (0.213)*** & 0.820 (0.087)*** \\
\textit{Activity (Out-links)} &  &  \\
Influential users & -1.391 (0.78)† & N/A \\
Leaders & -0.513 (0.783) & 1.250 (0.209)*** \\
Organizations (H3) & 1.351 (0.33)*** & 0.623 (0.206)** \\
\textit{Popularity   (In-links)} &  &  \\
Influential users & 0.842 (0.121)*** & N/A \\
Leaders & 0.199 (0.21) & 0.195 (0.075)** \\
Organizations (H4) & 0.702 (0.096)*** & 0.847 (0.077)*** \\
\textbf{Control variables} &  &  \\
Users’ popularity & -4.215 (0.135)*** & -6.363 (0.276)*** \\
Users’ activity & 2.831 (0.145)*** & 3.043 (0.186)*** \\
Sender’s number of followers & 0.284 (0.075)*** & -0.371 (0.081)*** \\
Receiver’s number of followers & 0.014 (0.023) & 0.024 (0.033) \\ \midrule
AIC & -91,157 & -133,061 \\
BIC & -91,003 & -132,940 \\ \bottomrule
\end{tabular}%
}
\caption{ERGM results for each connective network. Significance codes: *** $p < 0.001$, **, $p < 0.01$, * $p < 0.05$, † $p < 0.10$.}
\label{tab:results}
\end{table}

H2 states that crowd-enabled networks are more likely to form through triadic closure. The triadic closure term was positive and significant in the crowd-enabled network's ERGM ($\beta = 2.722, OR = 15.21, p<0.001$), which means that messages coming from direct contacts and sources were 15 times more likely to occur than chance. Therefore, H2 is supported.

H3 states that organizations are more likely to spread messages in mobilization networks. Both ERGMs confirm these hypotheses: the organizations' activity term was positive and significant in the crowd-enabled ERGM ($\beta = 1.351, OR = 3.861, p<0.001$) and in the organizationally-enabled ERGM ($\beta = 0.623, OR = 1.864, p<0.01$). These results show that organizations were almost four times more likely than individuals to retweet in the crowd-enabled network, and 1.86 times in the organizationally-enabled network. Thus, H3 is supported.

Finally, H4 states that messages published by organizations are more likely than messages published by individuals to be spread during the formation of mobilization networks. The organizations' popularity term was positive and significant in the crowd-enabled ERGM ($\beta = 0.702, OR = 2.017, p<0.001$) and in the organizationally-enabled ERGM ($\beta = 0.847, OR = 2.332, p<0.001$). This result indicates that organizations were more likely to be retweeted than individuals in both hashtags: two times in crowd-enabled network and 2.33 times in the organizationally-enabled network. Thus, H4 is supported.

The results also indicate that the crowd-enabled network was more hierarchical than the connective action posits. In both ERGMs, the users' popularity terms of the crowd-enabled ($\beta = -4.215, p<0.001$) and organizationally-enabled ($\beta = -6.363, p<0.001$) networks demonstrated that most attention was concentrated among a small number of accounts. Furthermore, we did not find evidence that users with many followers were retweeted more often. However, the users who were more likely to retweet others in the crowd-enabled network had a high number of followers ($\beta = 0.284, p<0.001$). In the organizationally-enabled network, users with a small number of followers were more likely to retweet ($\beta = -0.371, p<0.001$). Finally, influential users were 2.3 times more likely to be retweeted than ordinary users ($\beta = 0.842, OR = 2.321, p<0.001$), but influential users were 75\% less likely to retweet other users than ordinary users ($\beta = -1.391, OR = 0.248, p<0.10$). In the organizationally-enabled network, leaders were almost 3.5 times more likely to retweet others than ordinary users ($\beta = 1.250, OR = 3.489, p<0.001$) and 1.2 times to be retweeted ($\beta = 0.195, OR = 1.215, p<0.01$).

\section{Discussion}
This study analyzes the communication mechanisms and the role of organizations and individuals that shape the formation of digitally-enabled mobilization networks. 

Informed by the logic of connective action, we hypothesized that organizationally-enabled networks are more likely to form through hierarchical cascades than by chance (H1). Our findings indicate that organizations' high degree of control over personal action frames and strategic message dissemination fostered the formation of a hierarchical network. \#YoMarchoEl28's network was profoundly reliant on the connections and efforts of organization-managed accounts, reflecting how traditional organizations manage and deploy several resources to achieve their objectives \cite{McCarthy1977}. In contrast, ordinary users were less inclined to retweet messages within this network since the hashtag's personal action frame was narrowed to marching, providing fewer paths to adopt, re-signify, and perform these frames. Consequently, the network's formation was heavily driven by organizational accounts as the central information sources, resulting in a structure characterized by more hierarchical cascades than expected.  

Second, we posited that crowd-enabled networks are more likely to form through triadic closure than by chance (H2). Our analysis revealed that the \#FuerzaEstudiantes's network predominantly emerged through the retweets of shared and direct connections on Twitter. The endless possibilities to transform personal action frames allowed ordinary users to adopt and share them with their own connections. From identifying with the cause to providing solidarity to the students protesting, ordinary users had more motivation to retweet directly the accounts that started these messages. Additionally, the repressive police actions during this period allowed this frame to accelerate its diffusion, demonstrating the importance of social context and media events in driving diffusion. Retweeting functioned as a key mechanism to amplify the message from a cause that was in the periphery of the networks, resulting in more triadic closure structures. 

The results also show that organizations were crucial for the formation of these mobilization networks since they shared more content from others (H3) and were retweeted more frequently (H4) than ordinary users. Consistent with prior research, organizations remain instrumental to both crowd-enabled and organizationally-enabled mobilizations \cite{Gerbaudo2017,Karpf2016}. Acting as informational hubs among activists, organizations continue to be critical to catapult action frames to public spheres \cite{Friedlander2018}. While organizations may appear to play a role in the background within crowd-enabled networks, our results suggest that they and their leaders can facilitate connections of less connected users, thereby expanding the visibility of personal action frames. A potential explanation is that organizations function as hidden influencers within these networks \cite{GonzalezBailon2013,banos2013role}, taking advantage of their resource management capacity, coordinated actions, and dormant networks that reach thousands of disconnected individuals \cite{Pavan2017}.

Our study also highlights the importance of influential users and leaders, who were highly retweeted on crowd-enabled and organizationally-enabled networks, respectively. This finding underscores the differences in power and status among individuals and organizations, which are not highly described in connective action. Consistent with previous research, the presence of influential users––whether they are prominent public figures or highly connected users––brings greater visibility to social issues \cite{GonzalezBailon2013}. Influential users and leaders participating in these collective efforts enable these thousands of small networks to connect and establish a common discourse, which is ultimately facilitated by a highly personalized and inclusive action frame. Although influential users and leaders cannot claim ownership of these personal action frames, they facilitate the expansion of these frames among different groups on digital media and, ultimately, enable what Bennet and Segerberg described as a ``network of networks.'' Consequently, many of these online mobilizations are likely to disseminate quickly without recognizable faces or traditional organizational structures. 

Lastly, the two case studies discussed in this paper illustrate the diverse actions and roles undertaken by organizations during the 2011 Chilean Student Movement. Unlike many international protests that emerged from social media platforms \cite{Castells2011}, the Chilean movement demonstrated the significant role of formal and emergent organizations within both crowd-enabled and organizationally-enabled networks. The movement's effectiveness stemmed from an ecosystem promoting both kinds of networks. As shown, student organizations consistently coordinated protests and disseminated their frames on Twitter, while activists leveraged these networks to amplify their personal frames. This dynamic was further reinforced by the heightened political debates and contingencies that fostered sustained organizational activity from 2011 onward, leading to massive social protests in 2019 and the election of many student leaders in the 2022 national government. Therefore, the 2011 Chilean Student Movement underscores the importance of hybrid organizational structures in social movements, where formal and informal networks converge to sustain mobilization and amplify diverse voices in contentious political landscapes.

\subsection{Implications}
Our findings have important implications for understanding and shaping digital mobilization networks. Political implications highlight how both organizations and ordinary users contribute to shaping public discourse through distinct strategies. While organizations benefit from structured, coordinated actions and the presence of prominent leaders, ordinary users leverage their social networks and ability to create engaging content with their personal frames. As such, ordinary users are encouraged to focus on their existing relationships to spread messages, rather than mimicking the structured approaches used by organizations. As social media platforms facilitate the spread of these highly-personalizable messages, ranging from text to images and videos, ordinary users have the advantage of connecting directly with friends and social contacts to promote their mobilization groups' goals. Similarly, organizations and leaders should prioritize broad dissemination strategies (e.g., broadcasting) as well as actively listening and amplifying messages emerging from grassroots activists within their mobilization networks. 

Practical implications offer insights into how organizations and users can strategically optimize their efforts to spread social movements on social media platforms. A key factor is the role of accounts that act as hubs, bridging influential users and less connected users. Our results suggest that distributing messages and frames relies on bottom-up mechanisms, where certain accounts can bring visibility to messages posted by users on the periphery of their mobilization networks. Through highly connected accounts, these hub accounts can spread these mobilization messages to the rest of the network. Additionally, our study underscores the impact of social and political contexts---such as repression and protests---in shaping the trajectory of mobilization networks. Organizations, leaders, and activists often employ these contentious events to frame their messages in ways that resonate more deeply with the rest of the network. Ultimately, understanding these diffusion mechanisms will help both activists and leaders design more effective strategies to foster civic engagement, enhance political participation, and sustain collective action.

\subsection{Limitations and Future Work}
Our study is not exempt from limitations. First, more case studies are necessary to generalize our results to other social movements and social media platforms. This research examined two mobilization networks from Twitter that require more attention, especially in academic venues and outlets from the Global North. However, we acknowledge this case is not recent, and examining new case studies is necessary to demonstrate that these hypotheses have been substantiated. Second, the role of informal and institutionalized organizations should also be explored in more detail. We acknowledge that organizations differ in reputation and size, and people may trust them in different ways. These nuances should be explored in future studies. 

Third, content analysis can also provide better insights into users' motivations and goals with each interaction, such as the level of emotion involved in the frame. We did not analyze the content of each retweet, and incorporating text analysis could provide deeper insights into how users and organizations engage in these mobilization networks. Fourth, the sample size is relatively small compared to other research studies using Twitter data. We focused on 1,000 retweets to balance a substantial dataset and the feasibility of running ERGM models. Although we tested our models larger numbers of retweets, including additional edges increased the number of triangles in these networks, which introduced multicollinearity issues in the ERGMs. Extensions of this work should employ advanced machine learning models as well as traditional social network analysis techniques. Moreover, varying data collection thresholds should be explored to better understand the dynamics of these mobilization networks, such as adjusting the time frame (e.g., number of days, hours) and the number of retweets analyzed. Fifth, the dualistic vision of individuals and organizations may be simplistic and exclude intermediate organizational structures, such as small or emergent groups that appear on social media platforms. Future work should include more granular definitions and analyses to understand their differences.

Lastly, we did not examine the algorithmic components that mediated these retweet networks. In the early 2010s, Twitter timelines were based on a chronological order and algorithms did not mediate users' timelines significantly. This scenario is completely different in the 2020s, with multiple social media platforms highly mediated by recommendation algorithms and with a high presence of bots. Thus, the reliance on Twitter data from 2011 limits the applicability of these findings to current contexts. More recent mobilization networks may depart from the original concepts of connective actions, as algorithms now play a fundamental role in shaping the structure of these information networks. Future work should address the impact of algorithms, bots, and new social media platforms on the logic of connective action and their networks. 

\section{Conclusion}
The purpose of this study is to examine the influence of individuals' and organizations' communication actions and roles on the formation of digitally-enabled mobilization networks. By examining two mobilization networks of the 2011 Chilean Student Movement, we expand the logic of connective action by examining the structural signatures that characterize these mobilization networks: organizationally-enabled networks are more likely to spread through hierarchical structures than chance, and crowd-enabled networks will present more triadic closure structures than chance. Moreover, our results demonstrate that organizations are still relevant for both kinds of networks: they are more likely to share mobilization messages than individuals, and mobilization messages are more likely to be spread through organizations. 

Given the ongoing discussions of mobilizations enabled by social media platforms, our analysis shows that organizations continue performing a germane role in expanding and consolidating mobilizations. It will be important to consider how different communication mechanisms are likely to be executed in organizationally-enabled and crowd-enabled networks, assessing their different purposes and effectiveness given the leadership, organizational structures, and contingency. Most importantly, organizations continue to be relevant in the formation and consolidation of protest networks in online spaces, as they can be part of the front line or be articulators among several online activists. We hope this study contributes to shed light on the emergence of contemporary mobilizations. 

\begin{acks}
This work was supported by the University of Notre Dame's Democracy Catalyst Funds. We also thank the anonymous reviewers for their feedback and suggestions.
\end{acks}

\bibliographystyle{ACM-Reference-Format}
\bibliography{sample-base}

\afterpage{\clearpage}

\appendix

\section{Appendix}
\setcounter{figure}{0}
\counterwithin{figure}{section}

\begin{figure}[!htb]
    \centering
    \includegraphics[width=0.9\linewidth]{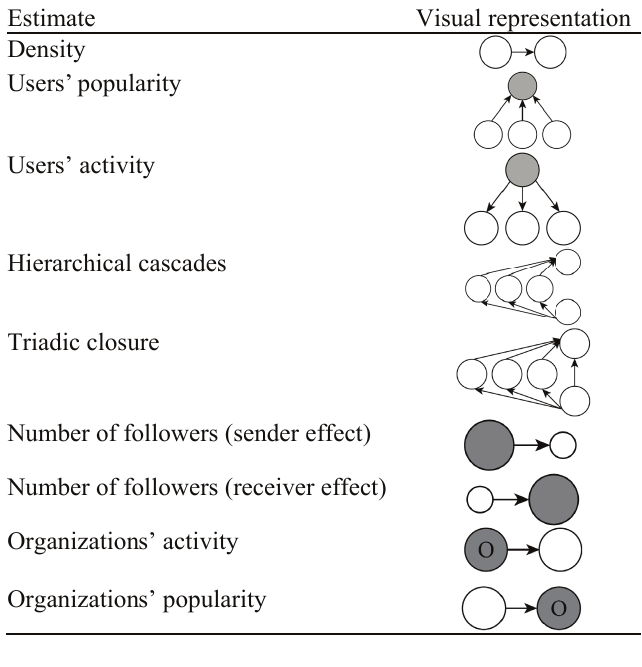}
    \caption{ERGM terms and visual representations}
    \label{fig:ergm-terms}
\end{figure}

\begin{figure}[!htb]
    \centering
    \includegraphics[width=\linewidth]{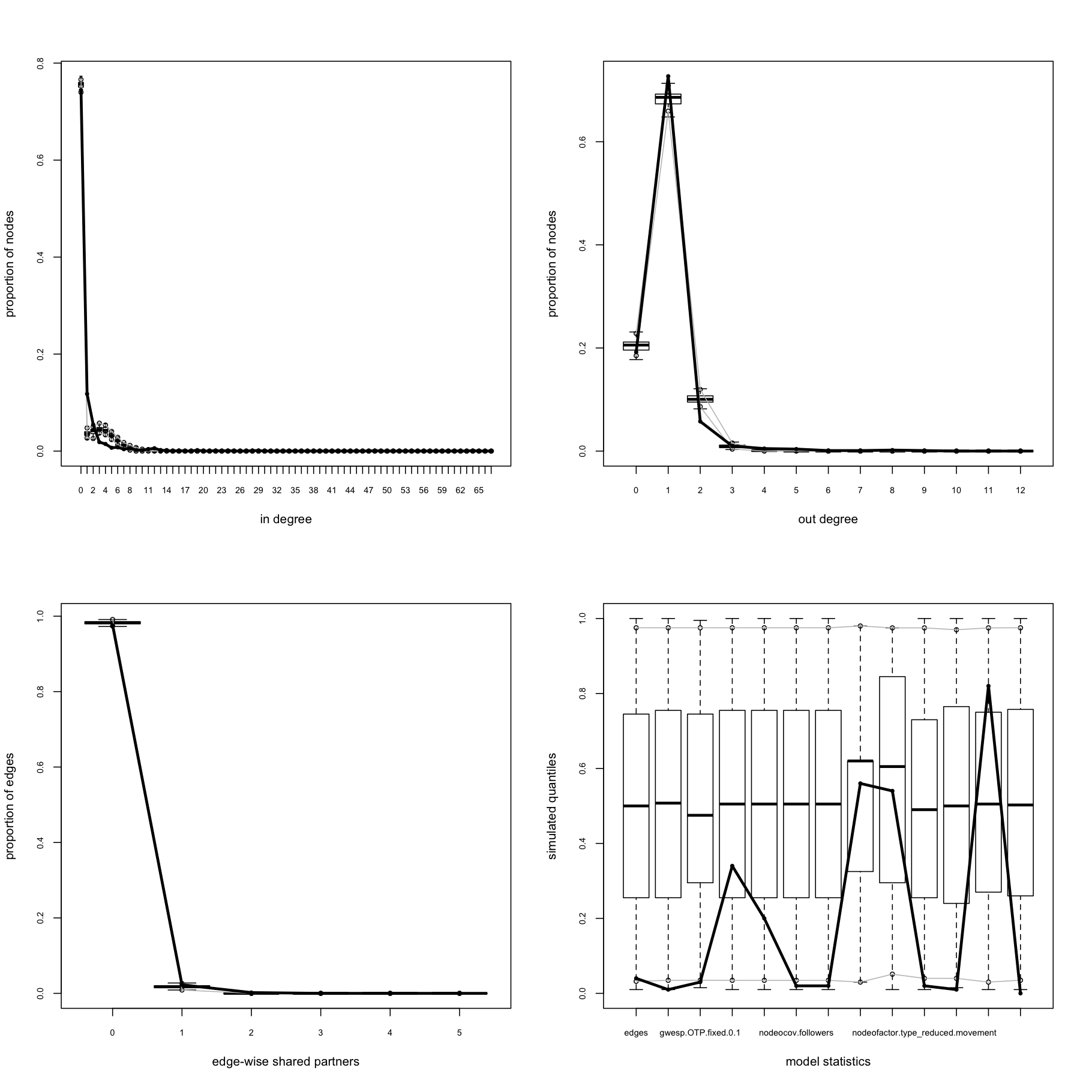}
    \caption{Goodness of Fit of the model statistics for the \#FuerzaEstudiantes ERGM.}
    \label{fig:gof-ergm1}
\end{figure}

\begin{figure}[!htb]
    \centering
    \includegraphics[width=\linewidth]{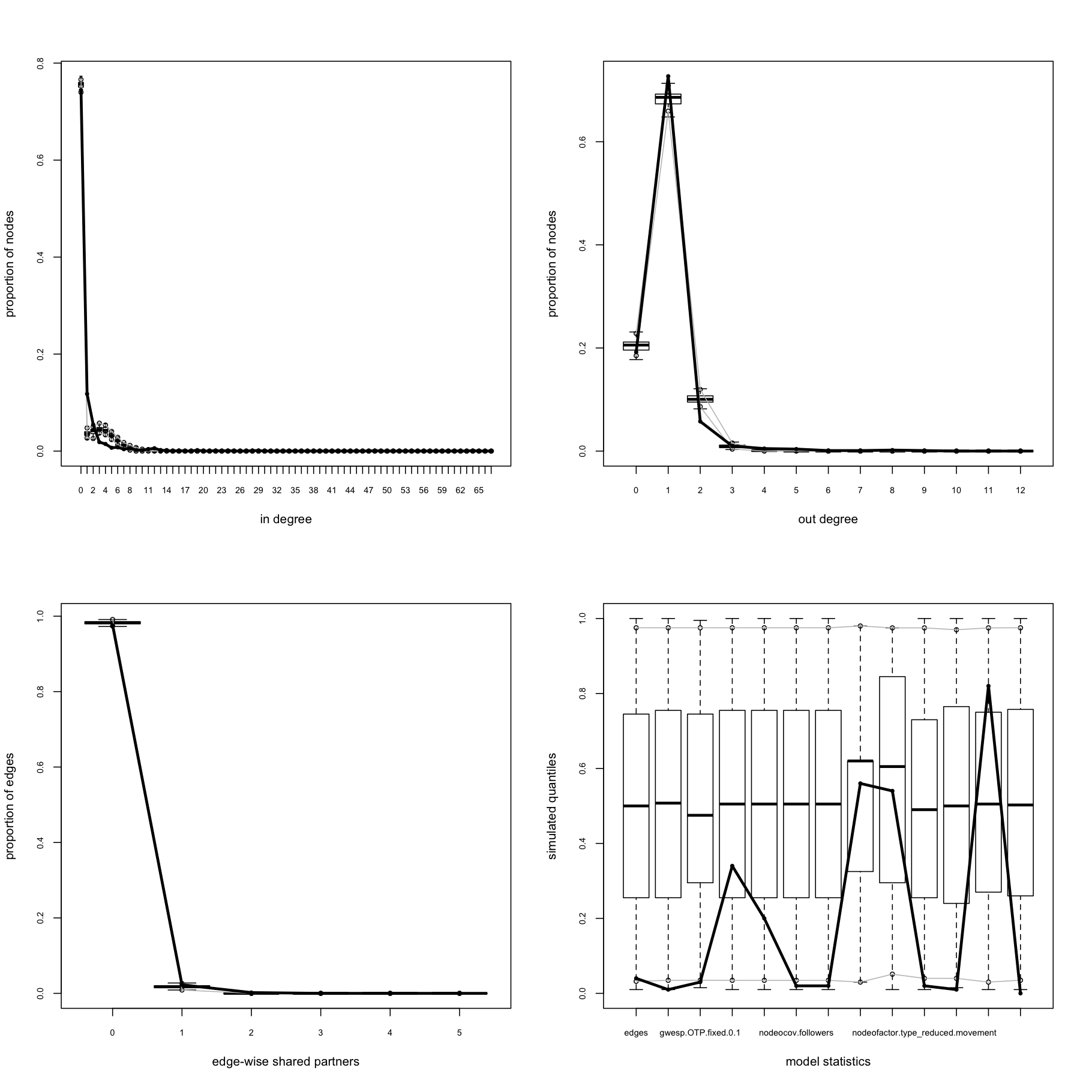}
    \caption{Goodness of Fit of the model statistics for the \#YoMarchoEl28 ERGM.}
    \label{fig:gof-ergm2}
\end{figure}

\end{document}